\documentstyle[12pt]{article}
\begin{document}

\begin{center}

{\Large {\bf THEORY OF DIELECTRIC ABSORPTION}}

\medskip

{\Large {\bf LINE SHAPE IN DIELECTRICS AND}}

\medskip

{\Large {\bf FERROELECTRICS}}

\bigskip

 M.D.~GLINCHUK, I.V.~KONDAKOVA

Institute for Material Sciences\\
National Academy of Sciences of Ukraine\\
Krzhizhanovskogo str. 3, 252180 Kiev, Ukraine

\end{center}

\medskip

\begin{abstract}
\noindent
Theory of dielectric spectroscopy spectra developed allowing for homogeneous and
inhomogeneous broadening mechanisms contributions. We calculated for the first
time dielectric absorption line form for an oscillator with memory and Debye
type relaxators with distribution respectively of resonant frequency and
relaxation time due to random electric field distribution. The conditions of
hole burning observation are discussed. It was shown that the investigation of
the inhomogeneously broadened line shape and width, hole burning can give
valuable information about the distribution of random field and relaxation time.
The latter seems to be extremely important for the disordered systems like
supercooled liquids and relaxator ferroelectrics.
\end{abstract}

distribution function, saturation.
%\pacs{}

\section{Introduction}

The dielectric spectroscopy is known to be one of the most utilized
method for dielectric and ferroelectric materials investigations (see
e.g.$^{[1]}$ and references therein). Physical information was obtained
mainly via analysis of the frequency, temperature and other external
parameters dependencies of the dielectric response maxima position.
Analysis of the dielectric spectra shape was practically absent.
Recently the observation of the spectral hole burning in the dielectric
response of supercooled liquids$^{[2]}$ had undoubtedly shown that
similarly to the magnetic resonance spectroscopy spectra investigated in
many details (see e.q.$^{[3,4]}$ ) the mechanisms of the dielectric
spectra broadening have to be divided into homogeneous and inhomogeneous
ones. This division is related to physical picture in what the
fluctuations of the local field "felt" by the dipoles can be divided
into the fast and slow comparing with the dipole relaxation. In such
picture the fast and slow fluctuations are the sources of homogeneous
and inhomogeneous broadening. In the ordinary dielectric and
ferroelectric dipole-lattice relaxation and dipole-dipole interaction
between the same dipoles can be the examples of homogeneous  broadening
mechanisms, whereas the interactions between the dipoles of the
different kinds as well as the local electric field inducing by the
impurities and the lattice imperfections are the examples of the
inhomogeneous broadening mechanisms. The disordered ferroelectrics,
polymers, supercooled liquids etc. with the strong internal random
electric field are known to have the non-Debye dielectric response
related to the relaxation time distribution in broad frequency region.
This behaviour is the consequence of the distribution for the barrier in
the dipole many-well potential that induced by the random fields
influence$^{[5]}$ . Therefore in disordered systems dipole-lattice
relaxation can be the source of the inhomogeneous dielectric response
broadening.

  In the present paper we calculated for the first time the dielectric
absorption line shape for an oscillator with memory for  the homogeneous
broadening mechanism with Gaussian distribution of its resonance frequency for
inhomogeneous broadening mechanisms. The distribution of relaxation time as
inhomogeneous broadening mechanism as well as the conditions for the hole
burning in the dielectric spectra are discussed.

\section{Theory of inhomogeneously broadened \newline
dielectric absorption}

2.1 \underline{Dielectrics Absorption}

\noindent
The intensity of dielectric absorption is known to be proportional to
  the number of electric dipoles which can be oriented by a.c. electric
  field whereas dipole-dipole interaction and relaxation processes tend
  to restore the thermal equilibrium state of the system. If the
  relaxation processes are slower than the rate of a.c. field absorption
  the number of the dipoles oriented along a.c. field increase and so
  the absorption decreases i.e. we are faced with the saturation of the
  response. Let us begin with the simplest case without saturation
  when the relaxation processes are fast enough. In such a case they
  are the same in all the points a main contribution to the homogeneous
  broadening of dielectric spectra. In the systems with local random
  electric field the dipoles in different points "feel"  the frequencies
  of lattice vibrations related to the distribution of the oscillator
  resonance frequencies.  Because it has to be $\Delta n_{i}$ dipoles
  ($n=\sum _{i} \delta n_{i}$ - the whole number of the dipoles)
  which "feel" the same field and so the same oscillator frequency
$\omega _{i}$ these dipoles can be
considered as a dipole packet (like spin-packet in magnetic system).
This packet absorption can be described as homogeneously broadened line
 $\phi (\omega -\omega _{i})$  with the width $\Delta =\frac{1}{\tau }$
and the intensity defined by the ratio $\Delta n_{i}/n$. In the limit
of continuous distribution of the resonance frequencies given by
$G(\omega _{i}-\Omega _{0})$ with the width $\Delta ^{*}$  we arrive
to inhomogeneously broadened spectra at the condition $\Delta
_{*}\gg\Delta $. In such a case the absorption line shape can be
represented as a convolution of the homogeneously broadened
dipole-packets with resonance frequency distribution:

\begin{equation}
g(\omega -\Omega  _{0})=\int_{-\infty }^{\infty }G(\omega _{i}-\Omega
_{0})\phi (\omega -\omega _{i})d\omega _{i}
 \label{1}
 \end{equation}
 The detail calculations were performed for oscillator
with memory defined by the decay function of the form$^{[6]}$
\begin{equation}
\alpha (t)=\gamma e^{-t/\tau } \sin \omega ^{'}t
\label{2}
\end{equation}
The Fourie-transform gives dielectric absorption as
\begin{equation}
\varepsilon ^{''}(\omega ,\omega _{0} )= \frac{4\gamma \omega \omega
^{'}} {(\omega ^{2}-\omega ^{2}_{0})^{2} +\frac{4\omega ^{2}}{\tau
^{2}}}
\label{3}
\end{equation}
 where $\omega _{0}^{2}=\omega ^{'2} +\frac{1}{\tau ^{'2}}$.
If the distribution of resonance
frequencies can be represented in Gaussian form
\begin{equation}
G(\omega _{i}-\Omega _{0})=\frac{1}{\sqrt{2\pi }\sigma }\exp \frac{-(\omega
_{i}-\Omega _{0})^{2}}{2\sigma ^2}
\label{4}
\end{equation}
Eq.~(1) with respect to
$\phi (\omega -\omega _i)\equiv \varepsilon ^{''}(\omega ,\omega_0)$
yields the line represented in Figure 1 for $\sigma \gg \Delta $
(inhomogeneously broadened line).
One can see that it is the envelope of the dipole-packets lines
(one of them depicted by doted lines). As the matter of fact Eq.~(2) can
be approximated by Lorentzian centered at $\omega _{i}$ with full width
$4/\tau ^2$ at the half maximum for the case of small damping$^{[7]}$.
 One can see by comparing Figure 1 and Figure 2 calculated for such
Lorentzian shape of homogeneously broadened packets that such
approximation is not bad at all.

%%%%%%%%%%%%%%%%%%%%%%%%%%%%%%%%%%%%%%%%%%%%%%%%%%%%%%%%%%%%%%%%%%%%%%%%%%
\begin{center}
\input{fig1.in}
\end{center}

\noindent
Fig.~1:
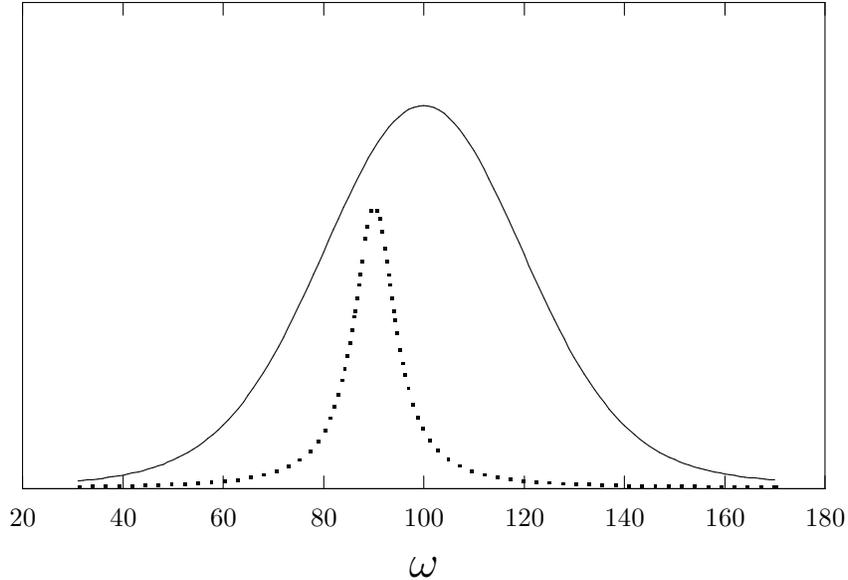
The inhomogeneously broadened line of dielectric absorption as a function
of frequency $\omega $ (arbitrary units).
Solid line is a convolution of the dipole packet lines (one of which is
depicted as the dotted curve, see Eq.~(3)) with Gaussian distribution
of the maximums of intensities.
%%%%%%%%%%%%%%%%%%%%%%%%%%%%%%%%%%%%%%%%%%%%%%%%%%%%%%%%%%%%%%%%%%%%%%%%%%%

\bigskip

\noindent
2.2 \underline{Disordered Ferroelectrics}

\noindent
In the case of strongly disordered systems like relaxor ferroelectrics
polymers, supercooled liquids the distribution of relaxation time up to
infinite one appears. In relaxor ferroelectrics the infinite time of
relaxation is related to the dipole glass state at $T<T_{g}$, where
$T_{g}$ is the freezing temperature. The distribution of the relaxation
time itself can be the source of inhomogeneous broadening  for
the dielectric absorption line. On other hand the long relaxation time
can lead to the saturation effects, i.e. to decreasing of the
absorption. Let us consider these phenomena in more the details.

Let us suppose that the dielectric response of the dipoles described by
the decay function of relaxation type, i.e.
\begin{equation}
\alpha (t)=\varepsilon _{s} \frac{1}{\tau } \exp{\frac{-t}{\tau }};
\varepsilon _{s}=\varepsilon _{0}-\varepsilon _{\infty}
\label{5}
\end{equation}
\noindent
can be written in the form of Debye law
\begin{equation}
\varepsilon ^{''}(\omega )=\frac{\varepsilon _{s}\omega \tau }{1+\omega
^{2}\tau ^{2}}
\label{6}
 \end{equation}
  with Arrhenius law for relaxation time
  \begin{equation}
  \frac{1}{\tau }=\frac{1}{\tau _{0}}\exp{(-\frac{U}{T})}
  \label{7}
  \end{equation}
  where $U$ is the height of the barrier between equivalent potential
  wells of a dipole. The existence of strong enough local random
  electric field has to influence the form of a dipole potential and so
  the form of the barrier and its height, namely
  \begin{equation}
  U(x)=U_{0}(x)\pm eE_{loc}x
  \label{8}
  \end{equation}
  For rectangular barrier this results into the following form
  relaxation time$^{5}$
  \begin{equation}
  \frac{1}{\tau }=\frac{1}{\tau _{0}}\exp{(\frac{-U\pm dE_{loc}}{T})}
  \label{9}
  \end{equation}
  The final form of relaxation time dependence on $E$ can be obtained
  after averaging of Eq.(9) over possible orientation of the electric
  dipole $d$. For two-orientable dipole it yields$^{[5]}$
  \begin{equation}
  \langle \tau \rangle =\bar{\tau
  _{0}}\frac{ch(2d^{*}E/kT)}{ch(d^{*}E/kT});
  \bar{\tau _{0}}=\tau _{0}\exp{U/T}
  \label{10}
   \end{equation}
   In Eq.~(10) we substituted the product $dE_{loc}$ by $d^*E$, where
   $d^*$ is the effective dipole moment.

   The form of the dielectric absorption can be obtained by substituting
   Eq.~(10) for $\tau $ into Eq.~(6) with the subsequent averaging with
   the random field distribution function $f(E)$, i.e.
   \begin{equation}
   \varepsilon ^{''}=\varepsilon _{s}\int \frac{\omega \tau f(E) dE}
   {1+\omega ^{2}\tau _{0}^{2}\frac{ch(2d^{*}E/kT)}{ch(d^{*}E/kT}};
   \label{11}
   \end{equation}
   Eq.~(11) is equivalent to that with the relaxation time distribution
   $F(\tau )$
   \begin{equation}
   \varepsilon ^{''}=\varepsilon _{s}\int \frac{\omega \tau F(\tau )
   d\tau } {1+\omega ^{2}\tau ^{2}}
   \label{12}
   \end{equation}
   introduced phenomenologically for many disordered systems.

   One can see that because for Debye law the position of the dielectric
   absorption maximum is defined by the condition $\omega \tau =1$ the
   distribution of $\tau $ leads to $\omega $ distribution, i.e. the
   line becomes inhomogeneously broadened. Note, that some another
   mechanism, e.g. dipole-dipole interaction, can be the source of homogeneous
broadening. When the width of distribution $f(E)$ or $F(\tau )$ is much larger
than that of dipole packet  the form of $\varepsilon ^{''}$ will be defined by
$f(E)$ or $F(\tau )$ calculated early in many details for linear
$^{[8]}$ and nonlinear random field contribution$^{[5,9,10]}$.

\section{Discussion}

\noindent
3.1 \underline{Hole Burning}

\noindent
This phenomenon is a direct evidence in favor of inhomogeneous broadening of
the line.  It was widely applied in radio-$^{[3,7]}$ and optic-$^{[8]}$
spectroscopy for investigation of random field distribution function and
relaxation processes in the solids.  The hole burning in dielectric spectra was
observed for the fist time recently$^{[2]}$.  Theoretical analysis of this
phenomenon in dielectric spectra is very restricted.

   The physical background of
hole burning is the saturation of one or some dipole-packets with the help of
powerful a.c. electric field pulse with the frequency
$\omega _{p}=\Omega_0+\Delta $
what belongs to the dipole-packet frequencies i.e.
$\omega_i\simeq \omega _p$.

  The time condition of this pulse application and
recording of dielectric response has to satisfy the demand that
the dipole-lattice relaxation as well as the spectral diffusion over the line
envelope should not destroy the induced by a.c. pulse saturation of the
dipole-packet.  However partly the aforementioned processes of the system
equilibrium state restoration shall disorder the dipoles and so decrease the
hole on the dielectric spectra. In particular, the width of the hole $2\delta $
depends on the degree of the packets overlapping what defines the velocity of
a.c. pulse energy transfer over whole inhomogeneous line. For the case when
these processes are very slow in approximation of Lorenzian dipole-packet
shape
  \begin{equation}
  \delta \sim \Delta \sqrt{1+S_{i}}
  \label{13}
  \end{equation}
  Here $S_{i}\sim W_{i}\tau $ is the saturation factor, what has to be
large enough for hole burning, $W_{i}\sim E^{2}_{a.c.}\varepsilon
^{''}_{max}$ ($\omega =\omega _{i}$) is the probability of induced by a.c.
field reorientations of dipoles in $i^{th}$ packet. The possible hole shape in
dielectric spectra is schematically represented in Figure 2.  Such type of
hole was did observed recently in supercooled liquids$^{[2]}$.

\begin{center}

\input{fig2.in}

\end{center}

\noindent
Fig.~2:
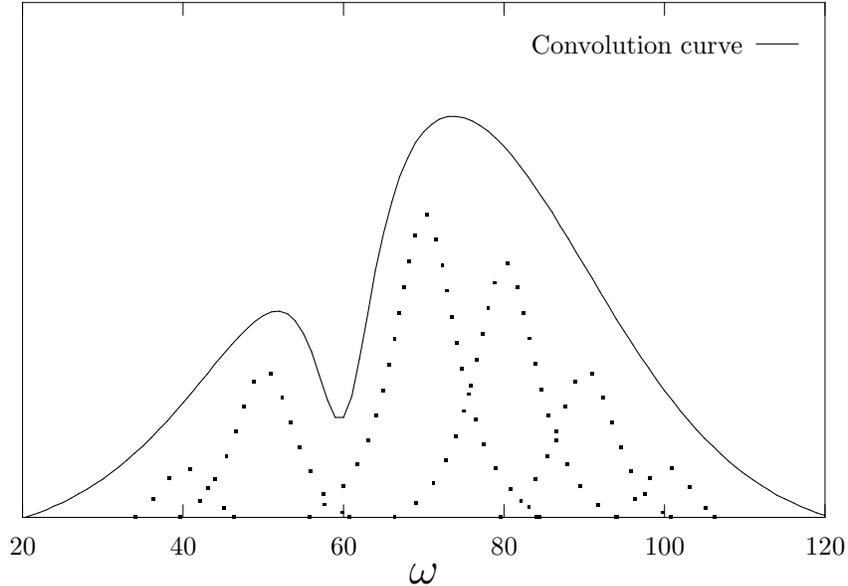
The shape of homogeneously broadened absorption spectrum with the
saturation of one of the dipole-packets at frequency $\omega _i$.
The shape of the dipole-packet is Lorenzian.
%%%%%%%%%%%%%%%%%%%%%%%%%%%%%%%%%%%%%%%%%%%%%%%%%%%%%%%%%%%%%%%%%%%%%%%%%%%

\bigskip

\noindent
3.2 \underline{Analysis of Dielectric Spectra and Hole-Burning}

\noindent
Let's briefly  discuss the physical information what can be obtained from
analysis of dielectric spectra shape and hole-burning.

  First of all the dielectric spectra
of the materials with non-Debye type of dielectric response have to be
inhomogeneously broadened. It is because such response is the manifestation of
relaxation time distribution, which has to be the source of the lines
inhomogeneous broadening. The hole burning can give a valuable information about
homogeneous and inhomogeneous broadening parameters. In particular,  if the
ratio of these broadening width
 $\frac{\Delta }{\Delta ^{*}}\ll 1$ the
$\varepsilon ^{''}$ shape has to be close to that of random field
distribution function. The latter quantity is known to be extremely important
for the physic of phase transitions in the disordered systems. In particular, it
defines the phase diagram of these systems$^{[8,11]}$. Note, that  under
the condition
 $\frac{\Delta }{\Delta ^{*}}\simeq 1$ or $\geq  2$  the
 hole burning is impossible because the spectra becomes
only slightly inhomogeneous or completely homogeneous. Therefore the
experimental investigation and the analysis of dielectric spectra shape and
hole burning can be the source of valuable information about physic of
dielectrics and disordered ferroelectrics in particular.

\pagebreak

\section*{References}

\noindent
1.
M.E.Lines, A.H.Glass, \underline{Principles and Application of
Ferroelectrics}\\
\underline{and Related Materials} (Clarendon Press, Oxford 1977).
\medskip

\noindent
2.
  B.Schiener, R.B\"ohner, A.Loidl, R.V.Chamberlin,
\underline{Scince},{\bf 274}, 752 (1996).
\medskip

\noindent
3.
 A.M.Portis, \underline{ Physical Review},
{\bf 91}, 1071 (1953).
\medskip

\noindent
4.
 P.L.Kuhs, M.S.Conradi, \underline{J.~Chem.~Phys.}
{\bf 77}, 1771 (1982).
\medskip

\noindent
5.
 V.A.Stephanovich, M.D.Glinchuk, B.Hilczer (in press).
\medskip

\noindent
6.
R.Blinc, Zeks, \underline{Soft Modes in Ferroelectrics and
  Antiferroelectrics} (North-Holland Publish.~Com.-Amsterdam, 1974).
\medskip

\noindent
7.
U.T.Hochli, K.Knorr, A.Loidl, \underline{Advance Phys.},
{\bf 39}, 405 (1990).
\medskip

\noindent
8.
 M.D.Glinchuk, V.A.Stephanovich,
 \underline{J.~Phys.:~Condens.~Matter}, {\bf 6},
6317 (1994).
\medskip

\noindent
9.
 M.D.Glinchuk, R.Farhi, V.A.Stephanovich, \underline{J.~Phys.:~
 Condens.}\\
 \underline{Matter}, {\bf 9}, 10237 (1997).
\medskip

\noindent
10.
M.D.Glinchuk, I.V.Kondakova,
\underline{Fizika Tverdogo Tela}, {\bf 40}, 340 (1998).
\medskip

\noindent
11.
  M.D.Glinchuk, R.Farhi, \underline{J.~Phys.:~Condens.~Matter}
{\bf 8}, 6985 (1996).

\end{document}